\documentclass[reprint,superscriptaddress,aps,prl,showemail,showpacs]{revtex4-1}

\usepackage{amsthm}
\usepackage{amsmath}
\usepackage{latexsym}
\usepackage{amsfonts}
\usepackage{amssymb}
\usepackage{color}
\usepackage{bbm,dsfont}
\usepackage{graphicx}
\usepackage{enumerate}
\usepackage{hyperref}
\usepackage{subfigure}
\usepackage{tikz}
\usepackage{placeins}
\usepackage[utf8]{inputenc}
\newcommand\opone{\leavevmode\hbox{\small1\kern-3.8pt\normalsize1}}

\DeclareUnicodeCharacter{00A0}{ }
 
\newtheorem{proposition?}{Proposition?}

\theoremstyle{definition}

\newcommand{\complex}{\mathbb C} 

\newcommand{\ket}[1]{|#1\rangle} 
\newcommand{\bra}[1]{\langle#1|} 
\newcommand{\Bket}[1]{\vert\vert#1\rangle} 
\newcommand{\Bbra}[1]{\langle#1\vert\vert} 
\newcommand{\kb}[2]{|#1\rangle\langle#2|} 
\newcommand{\Bkb}[2]{||#1\rangle\langle#2||} 

\newcommand{\id}{\mathbbm{1}} 

\newcommand{\vz}{\mathbf{z}} 
\newcommand{\vxi}{\boldsymbol{\xi}} 

\newcommand{\E}{\mathsf{E}}

\newcommand{\diff}{{\rm d}}
\newcommand{\norm}[1]{\vert\vert #1\vert\vert}
\newcommand{\abs}[1]{\vert #1 \vert}
\newcommand{\re}{{\rm Re}}

\newcommand{\mean}[1]{{\mathcal M}\left[ #1 \right]}

\begin{document}
\title{Parametrization and optimization of Gaussian non-Markovian
unravelings for open quantum dynamics}

\author{Nina Megier}
\email{nina.megier@tu-dresden.de}
\affiliation{Institut f{\"u}r Theoretische Physik, Technische Universit{\"a}t Dresden, 
D-01062,Dresden, Germany}

\author{Walter~T.~Strunz}
\affiliation{Institut f{\"u}r Theoretische Physik, Technische Universit{\"a}t Dresden, 
D-01062,Dresden, Germany}

\author{Carlos Viviescas}
\affiliation{Departamento de F{\'{\i}}sica, Universidad Nacional de Colombia, Carrera 30 No. 45-03, Bogota D.C., Colombia}

\author{Kimmo Luoma}
\affiliation{Institut f{\"u}r Theoretische Physik, Technische Universit{\"a}t Dresden, 
D-01062,Dresden, Germany}




\date{\today}


\begin{abstract}
We derive a family of Gaussian non-Markovian stochastic 
Schrödinger  equations for the dynamics of open quantum systems. 
The different unravelings correspond to different choices of
squeezed coherent states, reflecting different measurement schemes
on the environment. 
Consequently, we are able to give a single shot measurement interpretation 
for the stochastic states and microscopic expressions for
the noise correlations of the Gaussian process. By construction,
the reduced dynamics of the open system does not depend on 
the squeezing parameters. They determine the non-hermitian Gaussian correlation, 
a wide range of which are compatible with the Markov limit.
We demonstrate the versatility of our results for quantum information tasks 
in the non-Markovian regime.
In particular, by optimizing the squeezing parameters, we can tailor unravelings 
for optimal entanglement bounds or for 
environment-assisted entanglement protection.

\end{abstract}

\maketitle

\paragraph{Introduction.---}{Perhaps the most dramatic effect of the coupling of a quantum system
to an environment is the loss of quantum properties of its
state~\cite{breuer2}. Yet, decoherence seldom occurs in a simple
manner.  In the last decade, advances in experimental techniques made
it possible to observe non-Markovian dynamics in open quantum systems
as, for example, micromechanical~\cite{Groeblacher2015} and
optical~\cite{Liu2011} systems, highlighting the central part it plays
in preserving the coherent features of the system \cite{PhysRevLett.108.160402}. 
Non-Markovian dynamics 
has been proven to be essential for 
improvements in quantum metrology~\cite{Matsuzaki:2011bo,PhysRevLett.109.233601}, 
advances in quantum thermodynamics~\cite{Erez:2008ep} and optimal control
scenarios~\cite{Schmidt:2011bv}, in which the persistence of 
correlations such as entanglement  is crucial. 
The interplay of (non-Markovian) open system dynamics and time evolution
of quantum correlations is an active field of research~\cite{Aolita:2015fk}.  }

{General open quantum system dynamics can be approached 
from various perspectives. One can, as is most commonly done, use the
projection operator formalism~\cite{nakajima,zwanzig}, time local
master equations~\cite{BREUER200136} or hierarchical equations of
motion~\cite{tanimura,kreisbeck}, all of which describe the dynamics
on the level of the density matrix. Alternatively, a stochastic
description in terms of pure state unravelings 
(stochastic Schrödinger equations (SSEs))
is possible. Quantum jumps and
quantum state diffusion are then suitable methods both in the
Markovian~\cite{PhysRevLett.68.580,0305-4470-25-21-023} and
non-Markovian~\cite{PhysRevLett.100.180402,strunz_open_1999}
regimes. In particular, a complete parametrization of diffusive SSEs in the Markovian
regime is known~\cite{wiseman_complete_2001,chia_complete_2011}. 
Changing these parameters allows control over the noise correlations driving the
stochastic dynamics, which can be used to optimize the trajectories e.g. 
for entanglement
detection~\cite{viviescas_entanglement_2010,Guevara:2014eaa}. Moreover,
in the Markov case, a physical interpretation for the stochastic states 
can be given in terms of continuous monitoring of the environment of the open
system~\cite{Weber2014}.
}

{Recently, there have been similar efforts in the non-Markovian regime.
Di{\'o}si and Ferialdi~
\cite{diosi1,PhysRevLett.116.120402} have studied the structure of 
general non-Markovian Gaussian SSE
going beyond the standard non-Markovian quantum state 
diffusion (NMQSD) \cite{strunz_open_1999,walter3,
diosi_non-markovian_1997,Diosi98}.  
The same class of SSEs was then re-examined by
Budini from the perspective of its symmetries~\cite{budini}. However, 
a microscopic justification and derivation of general Gaussian non-Markovian 
SSEs is still lacking.}

{In this Letter we aim to fill this gap by providing a novel 
parametrization of the Gaussian noise 
correlations using squeezed states. We offer a 
single shot measurement interpretation for our family of non-Markovian
Gaussian unravelings. Due to the explicit parametrization and physical 
interpretation we are able to significantly improve entanglement bounds and 
perform environment-assisted entanglement protection in the non-Markovian regime.}

\paragraph{\it   {Open system model and general Gaussian unravelings}.---}
We {investigate} the dynamics of a system linearly coupled to a bosonic
bath. The Hamiltonian of the {total} system is given by $ H= H_S +
H_B + H_{SB}$, where $H_S$ is the Hamiltonian of the system, $H_B=
\sum_\lambda \omega_\lambda b_\lambda^\dagger b_\lambda$ is the bath
Hamiltonian, and $H_{SB}=\sum_\lambda g_\lambda(L b_\lambda^\dagger
+L^\dagger b_\lambda)$ describes their interaction. Here, $b_\lambda$
and $b_{\lambda}^{\dagger}$ are bosonic annihilation
and creation operators of the bath mode $\lambda$, with frequency 
$\omega_\lambda$, satisfying
$[b_{\lambda},b_{\lambda'}^{\dagger}]=\delta_{\lambda\lambda'}\id$. Furthermore $L$ 
is an arbitrary coupling operator acting on the system
and accounting for its interaction with all modes of the bath through
coupling amplitudes $g_\lambda$, {which, w.l.o.g., 
are chosen to be real}. {We switch to the 
interaction picture with respect to the
bath, in which the transformed Hamiltonian reads 
  $H_I(t) = H_S
    +\sum_\lambda g_\lambda(L b_\lambda^\dagger e^{i\omega_\lambda t} + 
    L^\dagger b_\lambda e^{-i\omega_\lambda t})$}.
{We assume}, for simplicity,
that the bath is initially at zero temperature,
$\rho_B(0)=\otimes_{\lambda} \kb{0_\lambda}{0_\lambda}$.

The key {ingredient} for our derivation of the {generalized Gaussian SSE
are \emph{Bargmann squeezed states}}.
For {mode $\lambda$}, these states are defined as {$\Bket{z_\lambda,\xi_\lambda}
  \equiv  R(z_\lambda,\xi_\lambda)\ket{0_\lambda}$},
{where} $R=\exp\left( z_{\lambda} b_{\lambda}^{\dagger} -
\frac{\xi_{\lambda}}{2} b_\lambda^{\dagger 2}\right)$, with  
$z_\lambda,\xi_\lambda\in\complex$, and $\abs{\xi_\lambda}<1$
\cite{dodonov}. Since
$R(z_\lambda,\xi_\lambda)$ is not unitary, the
$\Bket{z_\lambda,\xi_\lambda} $ are not normalized, yet the condition
$\abs{\xi_\lambda}<1$ guarantees that they are normalizable \footnote{The {Bargmann squeezed states} 
${\vert\vert}z,\xi{\rangle}$ used in the article 
differ from the usual ones ${\vert}z,\epsilon{\rangle} = \exp(z b^\dagger-z^* b)
\exp(\frac{\epsilon^*}{2} b^{2}-\frac{\epsilon}{2} b^{\dagger 2}){\vert}0{\rangle}$
with $z,\epsilon \in{\complex}$, $|\epsilon|<\infty $}. 
Most notably, the states $\Bket{z_\lambda,\xi_\lambda}$ are analytic both
in $z_\lambda$ and $\xi_{\lambda}$, and complete. In the multimode
bosonic environment, one can then write
\begin{equation}\label{eq:resolution_identity}
  \id=\int\diff^2\vz\, p_{\vxi}(\vz)\Bkb{\vz,\vxi}{\vz,\vxi}, 
\end{equation}
with $\Bket{\vz,\vxi}=\otimes_{\lambda}
\Bket{z_\lambda,\xi_\lambda}$, and measure $\diff^2\vz\,p_{\vxi}(\vz)
=\prod_\lambda\frac{\diff{\rm Re}z_\lambda\diff{\rm Im}z_\lambda}{\pi\sqrt{1-\abs{\xi_\lambda}^2}}
\exp\left(-\frac{\abs{z_\lambda}^2
-\frac{1}{2}(\xi_\lambda^*z_\lambda^2+\xi_\lambda z_\lambda^{*2})}{1-\abs{\xi_\lambda}^2}\right)$.

With the above completeness relation for squeezed coherent states at
hand, we now turn to the system dynamics. Using relation
\eqref{eq:resolution_identity}, a pure state $\ket{\Psi_t}$ of the
composite system, evolving according to Schr\"odinger equation
\begin{equation}
\label{eq:SGL}
\frac{\diff}{\diff t}\ket{\Psi_t}= -iH_I(t)\ket{\Psi_t},
\end{equation} 
can, at all times $t$, be expanded as
\begin{align}\label{eq:time_dependent_relative_states}
  \ket{\Psi_t}=\int\diff^2\vz\, p_{\vxi}(\vz)\ket{\psi_{\vxi^*}(\vz^*,t)}\Bket{\vz,\vxi},
\end{align}
where $\ket{\psi_{\vxi^*}(\vz^*,t)}\equiv\Bbra
{\vz,\vxi}\Psi_t\rangle$ is the \emph{unnormalized} state vector
of the system \emph{relative} to the environment squeezed 
coherent state $\Bket{\vz,\vxi}$.
{This approach is a generalization of non-Markovian quantum 
state diffusion with an additional freedom through 
squeezing parameters $\vxi$ \cite{walter4,Diosi98,strunz_open_1999}}.
We emphasize that $\ket{\psi_{\vxi^*}(\vz^*,t)}$ is an analytical function
of both $\vz^*$ and $\vxi^*$.
{Tracing over the bath, we find}
\begin{equation}\label{eq:mixstate}
\begin{split}
  \rho_S(t) &=\int\diff^2\vz\, p_{\vxi}(\vz)\kb{\psi_{\vxi^*}(\vz^*,t)}{\psi_{\vxi^*}(\vz^*,t)} \\
  & \equiv \mean{\kb{\psi_{\vxi^*}(\vz^*,t)}{\psi_t (\vz^*,\vxi^*)}}.
\end{split}
\end{equation}
That is, the reduced density operator of the system is obtained by
averaging over the unnormalized relative states with the Gaussian
probability density $p_{\vxi}(\vz)$; we denote this 
by $\mean{\cdot}$.  {Eq.~(\ref{eq:mixstate}) represents 
a family of unravelings of the open system dynamics parametrized 
by the squeezing parameters $\vxi$.}
{Moreover, decomposition
\eqref{eq:time_dependent_relative_states} allows for 
a single-shot measurement
interpretation of the unraveling; a topic to which we come back later
in the article.}

We are now in a position to state the {first} main result of
this article. Starting from the {Bargmann squeezed state}
representation of the {total} state, Eq.~(\ref{eq:time_dependent_relative_states}),  
we derive an {SSE} for the time
evolution of the relative states $\ket{\psi_{\vxi^*}(\vz^*,t)}$. 
{Combining Eqs. (\ref{eq:time_dependent_relative_states}) and (\ref{eq:SGL}),
and using the relations $b_\lambda^\dagger\Bket{z_\lambda,\xi_\lambda} =
\frac{\partial}{\partial z_\lambda}\Bket{z_\lambda,\xi_\lambda}$ and
$b_\lambda\Bket{z_\lambda,\xi_\lambda} =
\left(z_\lambda-\xi_\lambda\frac{\partial}{\partial
z_\lambda}\right)\Bket{z_\lambda,\xi_\lambda}$ \cite{dodonov,wuensche}, we are able to derive 
a closed linear non-Markovian SSE 
for the open system state $\ket{\psi_{\vxi^*}(\vz^*,t)}$},
\begin{equation}\label{eq:SSE_pre1}
\begin{split}
  \frac{\diff}{\diff t}\ket{\psi_{\vxi^*}(\vz^*,t)} &= -iH_S\ket{\psi_{\vxi^*}(\vz^*,t)} 
  + L z^*_{t}\ket{\psi_{\vxi^*}(\vz^*,t)}\\
 &- \int_{0}^t \! \diff s \left[\alpha(t,s)L^\dagger+\eta(t,s)L\right]  \frac{\delta}{\delta z^*_{s}}\ket{\psi_{\vxi^*}(\vz^*,t)}.
\end{split}
\end{equation}
{Here we use the chain rule $\frac{\partial}{\partial
z^*_{\lambda}}(\cdot) = \int \diff s\, \frac{\partial z^*_s}{\partial
z^*_{\lambda}}\frac{\delta}{\delta z^*_s}(\cdot)$, and introduce
the quantities}
\begin{subequations}
\begin{align}
\label{eq:noise}
z^*_{t}& \equiv -i\sum\limits_{\lambda}g_{\lambda} e^{i\omega_{\lambda}t}z^*_{\lambda},\\
\label{eq:hermitian_correlation}
\alpha(t,s)&\equiv \sum\limits_{\lambda}g_{\lambda}^2 e^{-i\omega_{\lambda}(t-s)},\\
\label{eq:non-hermitian_correlation}
\eta(t,s)&\equiv -\sum\limits_{\lambda}\xi^*_{\lambda}g^2_{\lambda} e^{i\omega_{\lambda}(t+s)}.
\end{align}
\end{subequations}

{Equation~(\ref{eq:mixstate}) shows that the reduced state dynamics 
is obtained by a Gaussian average over the solutions of Eq.~(\ref{eq:SSE_pre1}).
This amounts to regarding $z_t^*$ to be a Gaussian stochastic 
process, see Eq.~(\ref{eq:noise}). A simple calculation
gives $\mean{z_{t}^*}=\mean{z_{t}}=0$, and the
second order correlations
\begin{align}\label{eq:noise_correlations}
  \mean{z_{t}z_{s}^{*}}=\alpha(t,s),\quad
  \mean{z_{t}^*z_{s}^*}=\eta(t,s),
\end{align}
completely specifying the Gaussian process $z_t^*$.}

{The non-zero $\eta(t,s)$ correlation, determined 
by the squeezing parameter $\vxi$, is the new feature of our
generalized non-Markovian Gaussian SSE. The choice $\vxi=0$ leads to
$\eta(t,s)=0$, which is the standard NMQSD \cite{walter4,Diosi98,strunz_open_1999}.}

\paragraph{\it {Families of Gaussian unravelings, Markov limits and single shot measurements.}---}
{Since the partial 
trace over the environment is basis independent, it
is clear that the dynamics of the reduced state $\rho_S(t)$  cannot depend on 
the squeezing parameter $\vxi$ and 
therefore must be independent of  $\eta(t,s)$. However, different choices of 
$\vxi$, and thus $\eta(t,s)$, define different unravelings, Eq.~(\ref{eq:SSE_pre1}), with 
corresponding correlations, Eq.~(\ref{eq:noise_correlations}).}

{General non-Markovian Gaussian 
SSEs, similar to Eq.~(\ref{eq:SSE_pre1}), have been recently postulated
based on the properties of Gaussian processes and symmetry
properties of the system-environment interaction~\cite{diosi1,budini}. There 
the microscopic relations (\ref{eq:hermitian_correlation}), 
(\ref{eq:non-hermitian_correlation})
have to be replaced by  a positivity condition on the correlations,
which when applied here corresponds to the 
normalizability condition $\abs{\xi_\lambda}\leq 1$  mentioned earlier. 
The physics of our model determines $\alpha(t,s)$ to be a stationary 
correlation while $\eta(t,s)$ is a function of $t+s$. This is in contrast 
to the earlier approaches \cite{diosi1,budini}. In particular, 
in the Markov limit, when
$\alpha(t,s)\to \gamma \delta(t-s)$,  
the average dynamics follows
the Gorini-Kossakowski-Sudarshan-Lindblad (GKSL) master equation for
any choice of $\vxi$ in Eq.~(\ref{eq:non-hermitian_correlation}). 
We stress that this can be shown 
starting from Eq.~(\ref{eq:SSE_pre1}) without having to
resort to the microscopic origin. 
We can conclude that the Markov limit fixes solely the form of
the correlation $\alpha(t,s)$ leaving a wide range of choices for $\eta(t,s)$.}

Let us now return to state decomposition
\eqref{eq:time_dependent_relative_states} and consider the connection
it provides between our formalism and measurement theory.  A
\emph{single shot} measurement interpretation of the state
$\ket{\psi_{\vxi^*}(\vz^*,t)}$ in Eq.~\eqref{eq:SSE_pre1} can be offered: It is
the state of the system after a generalized measurement of the
bath with an outcome labeled by $\vz$ has been performed
at time $t$. This can be
made evident by noticing that the set of operators
$\E_{\vxi}(\vz)=\diff^2\vz\,p_{\vxi}(\vz)\Bkb{\vz,\vxi}{\vz,\vxi}$ is a
positive-operator valued measure (POVM). Then, from representation
\eqref{eq:time_dependent_relative_states} of state $\ket{\Psi_t}$, the
probability  of obtaining a measurement outcome
in the vicinity of $\vz$, when at a time $t>0$ a measurement of the
observable $\E_{\vxi}$ is done on the bath, is
\begin{equation}\label{eq:aposteriori_prob}
P_{\vxi}(\vz,t)\diff^2\vz= p_{\vxi}(\vz)\norm{\psi_{\vxi^*}(\vz^*,t)}^2\diff^2\vz. 
\end{equation}

{Freedom to choose $\vxi$ allows us to optimize the measurement on 
the environment for certain tasks. Next we will discuss two of them:
optimal bounds on entanglement dynamics \cite{walter0,
viviescas_entanglement_2010} and 
environment assisted error correction 
\cite{doi:10.1080/09500340308234541,PhysRevA.84.062314}.}

\paragraph{\it{SL-invariant entanglement measures.}---}
{We now address the problem of entanglement evolution
in open multipartite systems. First steps using a diffusive unraveling 
for the quantification of entanglement dynamics in non-Markovian open
system were taken in Ref.~\cite{walter0}.  With the new family of 
unravelings at hand, Eq.~(\ref{eq:SSE_pre1}), 
we are now in a position to tackle challenging tasks in 
quantum information dynamics in the non-Markovian regime.}

We consider the entanglement evolution in multipartite open systems in
which the subsystems do not interact among themselves, but one or more
of them may be coupled to its own local bosonic environment.  The
system consists  of $N$ subsystems, described by a Hilbert space
$\mathcal{H}_S=\mathcal{H}_1 \otimes \mathcal{H}_2 \otimes \cdots
\otimes \mathcal{H}_{N}$, each with arbitrary finite dimension. 
The system Hamiltonian $H_S$ is a sum of local Hamiltonians. 
Subsystem $k$
couples to its local bath through traceless operators $L_{k}$ with
real coupling amplitudes $g_{k,\lambda}$.

In order to quantify entanglement in this system we use special linear 
(SL)-invariant
multipartite measures of entanglement
$\mu_{\text{inv}}$~\cite{gour,Verstraete,Gheorghiu}. 
These are
polynomial measures defined by the following two properties: (i) They
are invariant under local linear transformations $U = U_1 \otimes U_2
\otimes \cdots \otimes U_N$, where $U_i$ acts on subsystem $i$ and
$\det U_i =1$, that is, $\mu_{\text{inv}}(U\psi) =
\mu_{\text{inv}}(\psi)$. (ii) They are homogeneous functions of degree
two for all $u\in \mathbb{C}$, i.e., $\mu_{\text{inv}}(u\psi) = |u|^2
\mu_{\text{inv}}(\psi)$. SL-invariant multipartite measures can be
used on mixed states by means of their convex roof extension
\begin{equation}
\label{eq:mixentanglement}
\mu_{\text{inv}}(\rho_S) = \min_{\{p_k,\psi_k\}}\sum_{k} p_k \mu_{\text{inv}}(\ket{\psi_k}),
\end{equation} 
where the minimum is taken aver all possible pure state decompositions
of $\rho_S$, i.e., $\rho_S=\sum_k p_k \ket{\psi_k}\bra{\psi_k}$
\cite{Eltschka}. The well known concurrence is the prime example 
of such an $SL$-invariant measure \cite{PhysRevLett.80.2245}.

Following Ref.~\cite{walter0}, for a system satisfying the above 
conditions, given an initial
normalized state $\ket{\tilde{\psi}(0)}$, a
scaling relation between the entanglement
$\mu_{\text{inv}}(\ket{\tilde{\psi}_{\vxi}(\vz,t)})$ of the normalized
relative state $\ket{\tilde{\psi}_{\vxi}(\vz,t)} =
\frac{\ket{\psi_{\vxi^*}(\vz^*,t)}}{\norm{\psi_{\vxi^*}(\vz^*,t)}}$  and the initial
entanglement in the system $\mu_{\text{inv}}(\ket{\tilde{\psi}(0)})$ can be 
established:
\begin{equation}\label{eq:scaling_relation}
  x_{\vxi}(\vz,t)\equiv
  \frac{\mu_{\text{inv}}(\ket{\tilde\psi_{\vxi}(\vz,t)})}{\mu_{\text{inv}}(\ket{\tilde\psi(0)})}=
    f_{\vxi}(\vz,t)\frac{P_{\vxi}(\vz,0)}{P_{\vxi}(\vz,t)}.
\end{equation}
The second equality follows from measurement outcome probabilities 
\eqref{eq:aposteriori_prob}, and the
details of the  scaling function $ f_{\vxi}(\vz,t)$ can be worked
out similar to Ref.~\cite{walter0}. 
Crucially, the new scaling relation now depends on the 
squeezing parameter $\vxi$.

\paragraph{\it {Optimal bounds on entanglement dynamics.}---}
Estimating and finding optimal bounds on multipartite entanglement
is a long-standing problem in entanglement theory \cite{Aolita:2015fk}. 
Based on Eq.~(\ref{eq:scaling_relation}), 
the freedom provided by the dependence of the scaling
function $ f_{\vxi}(\vz,t)$ on $\vxi$ allows us to look for tight upper
bounds on the entanglement $\mu_{\text{inv}}(\rho_S(t))$ of the 
reduced state of the system $\rho_S(t)$.
In
the framework of Markovian open quantum system dynamics diffusive equations
have been used to obtain optimal bounds 
\cite{viviescas_entanglement_2010,Guevara:2014eaa}. 
The new family of non-Markovian unravelings permits us to generalize
these results to the non-Markovian regime.
The pure state decomposition \eqref{eq:mixstate} provides an 
upper bound for the entanglement of the 
the open system state, we find   
\begin{align}\label{eq:inequality}
\frac{\mu_{\text{inv}}(\rho_S (t))}{\mu_{\text{inv}}(\ket{\tilde{\psi}(0)})}\leq \overline{x}_{\vxi}(t),
\end{align}
where $\overline{x}_{\vxi}(t)$ is the mean entanglement in the multipartite open
system. Here we use Eq.~\eqref{eq:scaling_relation}, that also leads to the  
expression $\overline{x}_{\vxi}(t)=\int \diff^2 \vz P_{\vxi}(\vz,0)
f_{\vxi}(\vz,t)$.

Remarkably, both the scaling relation
\eqref{eq:scaling_relation} and the upper bound \eqref{eq:inequality}
are independent of the {initial state as well as of the specific entanglement measure used, as long as it is SL-invariant.}

We demonstrate the significance of our findings with 
an example of non-Markovian  multipartite open
quantum system dynamics.  For
concreteness, let us assume that $M$ ($M\leq N$) of the subsystems are
qubits (two-level systems), each one of them coupled to its own local
dephasing bath via $L_k=L=\sigma_{z}$, ($k=1\dots M$), while the rest
of the $N-M$ subsystems remain isolated \footnote{
In Ref.~\cite{walter0} it was shown that
only the nature of the subsystems coupled to 
their respective  baths need to be
specified. The scaling relation, and therefore the entanglement
bound, is independent of the type of the constituents of the multipartite
open system which are not directly interacting with an environment.}. For
dephasing environments the scaling function becomes independent of
$\vz$ (see Ref.~\cite{walter0}) and the mean entanglement in the
system reduces to
\begin{equation}\label{boundsz}
\overline{x}_{\vxi}(t)=f_{\vxi}(t)=\prod_{k}^{M}\exp\left(-\frac{1}{2} \int\limits_0^t\diff s\, \gamma_{k}(s)\right),
\end{equation}
with time-dependent dephasing rates
\begin{equation}\label{eq:dacayrate}
\gamma_{k}(s) = 4\re \int_{0}^{s}\diff s'\,\left(\alpha_{k}(s,s')+\eta_{k}(s,s')\right). 
\end{equation} 

\begin{figure}
	\includegraphics*[scale=.58]{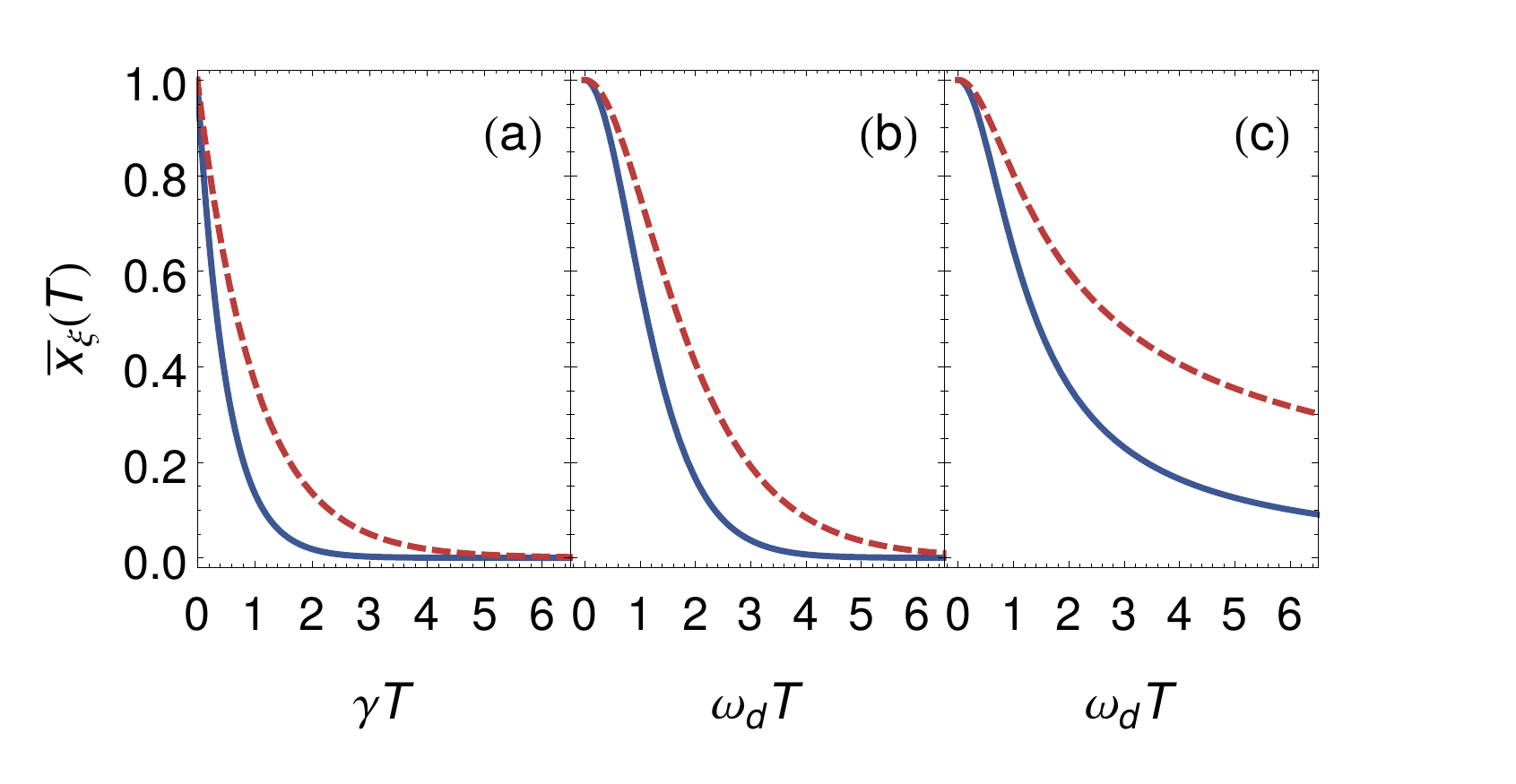}
	\caption{\label{fig:exact}(Color online) Mean entanglement
evolution in a $N$-partite open system, where only one 
subsystem of dimension two is exposed to a
dephasing channel. (a) Markov, (b) Ohmic, and (c) super-Ohmic
environment \cite{walter0}, where $\gamma$ is the 
Markov decay rate and $\omega_d$ is the 
cut-off frequency. In all cases the upper bound given by
$\overline{x}_{\vxi^{\text{opt}}}(T)$ (blue continuous line) coincide
with the exact dynamics of the reduced state  entanglement
(cf. Ref.~\cite{Guevara:2014eaa,anna}). For comparison we show the
bound obtained for $\overline{x}_{\vxi=0}(T)$ (red dashed line),
corresponding to the bound using the standard
NMQSD ($\eta=0$).}
\end{figure}

Clearly, any choice of $\eta_k$ provides, via \eqref{boundsz}, an
upper bound on the entanglement of the state of
the system $\rho_S(t)$. One can now ask for the optimal
choice $\eta_k^{\text{opt}}$ (and therefore $\vxi_k^{\text opt}$), which would yield the tightest
possible bound. Before going into the search for this optimal
unraveling a reminder on the meaning of our theory is due here. As a result of 
the single shot measurement interpretation of the system state, an
optimization of the mean entanglement $\overline{x}_{\vxi}$ at time $t=T$ must
target that specific time and may not be the optimal choice for a
different time $t\neq T$. With this in mind, we may now return to the
task of minimizing $\overline{x}$ in \eqref{boundsz}. We assume for simplicity that 
all local dephasing channels are identical so that 
$\gamma_k(t)=\gamma(t)$. For a given time 
$T$, $\overline{x}_{\vxi}(T)$ is minimal if 
the integral in Eq.~(\ref{boundsz}) 
is maximized for each bath mode $\lambda$.
A simple
calculation shows that the optimal value 
for $\overline{x}_{\vxi}(T)$ is obtained by setting the squeezing parameter
$\xi_\lambda^{\text{opt}}=-e^{i\omega_\lambda T}$~\footnote{Due to the
normalization constraint $|\xi_\lambda| < 1$, the optimal choice of the
squeezing parameter $\xi_\lambda^{\text{opt}}=-e^{i\omega_\lambda T}$ is to be
understood in the limiting sense.}, and
yields the upper bound
\begin{align}\label{eq:resiliance_factor}
\overline{x}_{\vxi^{\text opt}}(t)=\exp\left(-\frac{M}{2}
\int\limits_0^t\diff s\, \gamma^{\text{opt}}(s)\right).
\end{align}
Exact results on entanglement evolution in multipartite open systems
with non-Markovian dynamics are scarce \cite{Aolita:2015fk}, making it
difficult to asses how tight our bound really is. Yet, for the case of
two qubits with only one of them coupled to a dephasing channel, the
exact entanglement dynamics is given in \cite{Guevara:2014eaa} for 
Markovian and in \cite{anna} for non-Markovian dynamics,  
our bound exactly reproduces this entanglement 
evolution. 
Indeed, our bound is also exact for any $N$-partite open system, where only one 
subsystem of dimension two is exposed to an arbitrary dephasing channel~\cite{gour}. 
{In  Fig.~\ref{fig:exact} we show the entanglement dynamics,  
the optimal
entanglement bound (indistinguishable) and the previously obtained $\vxi=0$ bound 
for Markov-, Ohmic- and super-Ohmic 
dephasing environments.}

The multi channel result, Eq.~(\ref{eq:resiliance_factor}), coincides 
with the upper bound of the entanglement dynamics for a multipartite
mixed state given in 
Corollary 4 of Ref.~\cite{gour}.

\paragraph{{Environment-assisted entanglement protection.}---}
It has been shown that iff the dynamics of the 
open system is given by a random unitary channel, then 
there exists a protocol perfectly restoring the lost 
quantum information \cite{doi:10.1080/09500340308234541}. The error
correction scheme is conditional on the measurement performed on 
the system's quantum environment. Such a procedure has been
explicitly constructed for two qubit~\cite{PhysRevA.84.062314}
and for $N$-qubit~\cite{PhysRevA.88.022321} pure dephasing dynamics.

Our dephasing channel is of random unitary type. However, having 
restricted the measurement to Bargmann squeezed state POVMs we don't 
expect to recover the initial state but instead we aim to restore the 
initial entanglement.  

Indeed, by choosing $\xi_\lambda^{\text{restore}}=e^{i\omega_\lambda T}$, 
Eq.~(\ref{boundsz}) gives the bound $\overline{x}_{\vxi^{\text{restore}}}(T)=1$. 
This means that for any outcome $\vz$ 
of this optimal measurement at time $T$ the conditional state of the 
open system contains the initial amount of entanglement.

Remarkably, we are able to construct explicitly a measurement on 
a realistic quantum environment that realizes an environment-assisted 
entanglement protection scenario, generalizing earlier considerations on Markov open quantum
systems 
\cite{Vogelsberger:2010bu,Carvalho:1322471,Guevara:2014eaa}.

{Let us finally remark that with our SSE we 
are able to asses the
dynamics of entanglement without first solving the reduced
state dynamics~\cite{Guevara:2014eaa}.}

\paragraph{\it Conclusions.---} 
{
In the present Letter we derived a generalized Gaussian non-Markovian SSE by
expanding the environment in a  Bargmann squeezed state basis.
Each choice of the squeezing parameter $\vxi$ corresponds to a 
different unraveling and reflects a different measurement done on the environment
of the open quantum system. Thus our results also add to the discussion 
on the objectivity of collapse models in the non-Markovian regime
~\cite{PhysRevLett.108.220402,PhysRevLett.103.050403}.
Our microscopic approach 
leads to a stationary hermitian correlation $\alpha(t,s)$ and 
to a non-stationary non-hermitian correlation $\eta(t,s)$ for 
the Gaussian noise $z_t^*$. 
By construction, the reduced dynamics is independent of $\eta(t,s)$.
In the Markov limit we see that a wide range 
of different $\eta(t,s)$ are allowed, being compatible 
with the GKSL-dynamics.
We demonstrated the power of our family of unravelings for quantum information tasks 
in the non-Markovian regime.
In particular, for local quantum channels, by optimizing over 
the squeezing parameter $\vxi$, we can tailor the ensemble of 
relative states for optimal entanglement bounds or for 
environment assisted entanglement protection.

C.V. is thankful for the hospitality extended to him by the 
Institut für Theoretische Physik at Technische Universität Dresden.
\bibliography{SqueezedStateQSDBib}
\end{document}